# Multi-Faced Entanglement

Moses Fayngold

*Department of Physics, New Jersey Institute of Technology, Newark, NJ 07102*

Detailed analysis of behavior of spin-entangled particle pairs under arbitrary rotations in their Hilbert space has been performed. It shows a rich range of varieties (faces) of entanglement in different bases. Analytic criteria are obtained for the respective faces of an entangled state. The corresponding conditions generally depend on both the state itself and the chosen basis. The most important result is revealing a deep analogy between a spin-entangled electronic qubit pair and momentum-entangled photon pair. Both cases exhibit coherence transfer from individual particles to nonlocal state of the system. This analogy allows us to predict certain features of the interference patterns in spin-entangled qubit pairs.

# 1. Introduction.

The article presents a thorough analysis of behavior of spin-entangled systems, specifically – pairs of spin-(1/2) particles under rotations in their Hilbert space $\mathcal{H}$. Starting from a single particle (Sec.2), we formulate transformation rules for switching from one basis to another. Then we consider some characteristics of disentangled and entangled pairs in different bases and representations (Sec.3). In Sec.4, the transformation rules are applied to an entangled pair with opposite spin components. The analysis demonstrates entanglement as a very "flexible" physical characteristic changing its "face" in different bases. Multi-faced nature of entanglement is also shown for a system with equal spin-components in its reference basis (Sec.5). Finally, the case with different bases used by different observers is discussed (Sec. 6). In all cases, the corresponding analytic criteria are obtained for the respective "faces" of an entangled state. The found conditions, formulated in boxed equations, generally depend on both – the state itself and the chosen basis. One of the most important results of the analysis is that all studied cases of entanglement exhibit coherence transfer from individual particles to nonlocal state of the whole system. Such effect has also been known for entangled photon systems. In this respect, the studied system demonstrates a deep analogy between a spin-entangled fermion pair and momentum-entangled boson pair. Coherence transfer turns out to be fundamental feature of any entangled composite state. This allows us to predict some experimentally observable characteristics in the nonlocal bi-particle interference similar to known interference patterns in the nonlocal bi-photon states.

All essential points of the article are summarized in Sec. 7.

## 2. Spin-state in different bases

As a prelude to delving into possible entangled states of a bipartite system, we review first a single qubit state as recorded in different bases. Here, we will represent qubit as the spin (1/2) eigenstates. In the $s_z$-basis with eigenstates $|\uparrow\rangle$ ("spin up") and $|\downarrow\rangle$ ("spin down"), an *arbitrary* spin state $|\mathbf{s}\rangle$ shown in Fig.1a as an arrow **s** in the Bloch sphere [1-4], is a superposition

$$|\mathbf{s}\rangle = a|\uparrow\rangle + be^{i\varphi}|\downarrow\rangle; \qquad a = \cos\frac{\theta}{2}, \quad b = \sin\frac{\theta}{2} \qquad (2.1)$$

Its "antipode" $|\bar{\mathbf{s}}\rangle$ will be represented by an arrow $\bar{\mathbf{s}} \equiv -\mathbf{s}$ with $\bar{\theta} = \pi - \theta, \; \bar{\varphi} = \varphi + \pi$ and is expressed in terms of $a$, $b$ as

$$|\bar{\mathbf{s}}\rangle = b|\uparrow\rangle - ae^{i\varphi}|\downarrow\rangle \qquad (2.2)$$

But the model representing $|\mathbf{s}\rangle$ as a vector **s** in our space $V$ like in Fig.1a is largely misleading. Spin state $|\mathbf{s}\rangle$ is a vector in $\mathcal{H}$ but not in $V$. A better visual model would be a *cone* with symmetry axis **s** and an *open angle* $2\xi_{s,i} = 2\mathrm{Arccos}\left(|s_i|/\sqrt{s(s+1)}\right)$ where $s_i$ is one of the eigenvalues of $\hat{s}$ in the $(\mathbf{s}, \bar{\mathbf{s}})$-basis (Fig. 1b) and $\xi_{s,i}$ is the angle between **s** and a generatrix of its cone. Any "vector image" of spin hereafter will actually be only symmetry axis of the respective cone. Whereas angles $\varphi, \theta$ are continuous variables, angle $\xi_{s,i}$ is quantized, with only one value $\xi_{s,i} \to \xi_s = \text{Arccos}(1/\sqrt{3})$ for spin 1/2. In other words, $\varphi, \theta$ determine orientation of the cone with "frozen" open angle $2\xi_{s,i}$. Angles $\varphi, \theta$ also determine the probability amplitudes in superposition (2.1). They can be monitored by varying the direction of magnetic field **B** in a Stern-Gerlach apparatus used for preparing state $|s\rangle$. So $\theta, \varphi$ in (2.1) may actually indicate the direction of **B**.

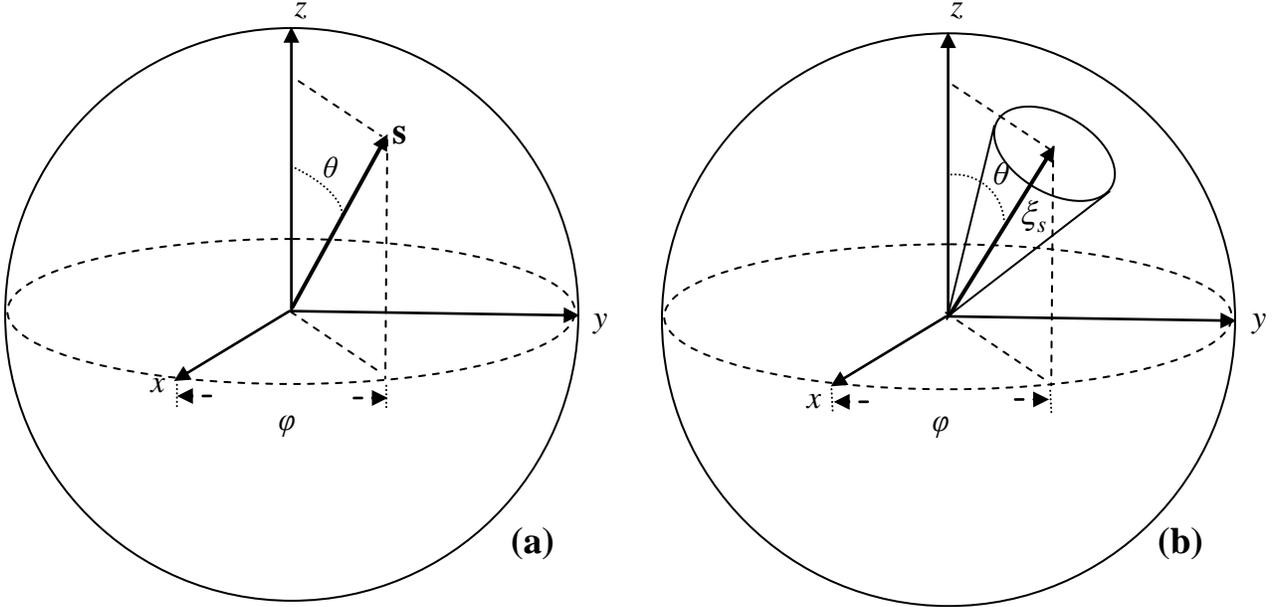

**Fig. 1**

Graphical representation of $|s\rangle$ using the Bloch sphere.

(**a**) State $|s\rangle$ is represented in physical space $V$ as a unit vector **s** with the polar angle $\theta$ and azimuth $\varphi$ ($(x, y, z)$ is the respective Cartesian triad).

(**b**) A more detailed representation (not to scale) is a cone with open angle $2\xi_s$ and symmetry axis along $\mathbf{s}(\theta, \varphi)$

We can abbreviate (2.1), (2.2) to a matrix form by treating the respective kets as C-numbers:

$$\begin{pmatrix} |s\rangle \\ |\bar{s}\rangle \end{pmatrix} = \mathcal{R}_\mathbf{s} \begin{pmatrix} |\uparrow\rangle \\ |\downarrow\rangle \end{pmatrix}, \qquad \mathcal{R}_\mathbf{s} \equiv \begin{pmatrix} a & be^{i\varphi} \\ b & -ae^{i\varphi} \end{pmatrix} \qquad (2.3)$$

In view of (2.1), the determinant of rotational matrix $\mathcal{R}_\mathbf{s}$ is $D(\mathcal{R}_\mathbf{s}) = -e^{i\varphi}$.

Suppose **s**, $\bar{\mathbf{s}}$ lie in the "equatorial" plane $\theta = \pi/2$, with $\varphi = 0$ for **s** and $\varphi = \pi$ for $\bar{\mathbf{s}}$. Then the pair **s**, $\bar{\mathbf{s}}$ forms the $s_x$-basis with eigenstates $|\odot\rangle, |\otimes\rangle$ (pointing respectively towards and away from us):

$$|\odot\rangle = \frac{1}{\sqrt{2}}(|\uparrow\rangle + |\downarrow\rangle); \quad |\otimes\rangle = \frac{1}{\sqrt{2}}(|\uparrow\rangle - |\downarrow\rangle) \qquad (2.4)$$

The inverse transformations:

$$|\uparrow\rangle = \frac{1}{\sqrt{2}}(|\odot\rangle + |\otimes\rangle); \quad |\downarrow\rangle = \frac{1}{\sqrt{2}}(|\odot\rangle - |\otimes\rangle) \qquad (2.5)$$

Setting in (2.1, 2) $\theta = \pi/2$, $\varphi = \pm\pi/2$ gives the $s_y$ basis $|\rightarrow\rangle$ ("spin-right") and $|\leftarrow\rangle$ "spin-left"):

$$|\rightarrow\rangle = \frac{1}{\sqrt{2}}(|\uparrow\rangle + i|\downarrow\rangle); \quad |\leftarrow\rangle = \frac{1}{\sqrt{2}}(|\uparrow\rangle - i|\downarrow\rangle) \qquad (2.6)$$

$$|\uparrow\rangle = \frac{1}{\sqrt{2}}(|\rightarrow\rangle + |\leftarrow\rangle); \quad |\downarrow\rangle = \frac{i}{\sqrt{2}}(|\rightarrow\rangle - |\leftarrow\rangle) \qquad (2.7)$$

If **s** is not in the equatorial plane, superposition (2.1) is not equally-weighted. For **s** shown in Fig.1 the outcomes $|\uparrow\rangle$ in the $s_z$–measurement are more probable than outcomes $|\downarrow\rangle$.

A basis formed by oppositely-directed vectors **s**, **s̄** in *V*, is orthonormal in $\mathcal{H}$,

$$\langle \mathbf{s}|\mathbf{s}\rangle = \langle \mathbf{\bar{s}}|\mathbf{\bar{s}}\rangle = 1, \quad \langle \mathbf{s}|\mathbf{\bar{s}}\rangle = 0 \qquad (2.8)$$

On the other hand, any two eigenstates from some *different* bases are non-orthogonal in $\mathcal{H}$ even if they reside in mutually perpendicular dimensions of *V*, like $|\uparrow\rangle, |\rightarrow\rangle$ or $|\odot\rangle, |\downarrow\rangle$, so that

$$\langle \rightarrow|\downarrow\rangle = e^{i\pi/2}\cos(\pi/4) = i/\sqrt{2}; \quad \langle \odot|\downarrow\rangle = \cos(\pi/4) = 1/\sqrt{2} \qquad (2.9)$$

The difference between $\mathcal{H}$ and *V* is also reflected by the fact that we have $\theta/2$ instead of $\theta$ in (2.1), (2.2), so *a*, while being von Neumann projection [5] of $|\mathbf{s}\rangle$ onto $|\uparrow\rangle$, is *not* geometrical projection of **s** onto *z* (let alone its different physical meaning as the probability amplitude).

Suppose we want to switch to a basis $S_\mathbf{e}$ along some direction **e** characterized by the polar angle $\chi$ and azimuth $\delta$. Let $|\mathbf{e}\rangle$ and $|\mathbf{\bar{e}}\rangle$ be the respective eigenstates. Then, similar to (2.1-3), we will have

$$\left.\begin{array}{l} |\mathbf{e}\rangle = m|\uparrow\rangle + ne^{i\delta}|\downarrow\rangle \\ |\mathbf{\bar{e}}\rangle = n|\uparrow\rangle - me^{i\delta}|\downarrow\rangle \end{array}\right\}, \quad m = \cos\frac{\chi}{2}, \ n = \sin\frac{\chi}{2} \qquad (2.10)$$

or

$$\begin{pmatrix} |\mathbf{e}\rangle \\ |\mathbf{\bar{e}}\rangle \end{pmatrix} = \mathcal{R}_\mathbf{e} \begin{pmatrix} |\uparrow\rangle \\ |\downarrow\rangle \end{pmatrix}, \quad \mathcal{R}_\mathbf{e} \equiv \begin{pmatrix} m & ne^{i\delta} \\ n & -me^{i\delta} \end{pmatrix}, \ D(\mathcal{R}_\mathbf{e}) = -e^{i\delta} \qquad (2.11)$$

The inverse transformation is

$$\begin{aligned}|\uparrow\rangle &= m|\mathbf{e}\rangle + n|\bar{\mathbf{e}}\rangle \\ |\downarrow\rangle &= e^{-i\delta}\left(n|\mathbf{e}\rangle - m|\bar{\mathbf{e}}\rangle\right)\end{aligned} \tag{2.12}$$

or

$$\begin{pmatrix}|\uparrow\rangle \\ |\downarrow\rangle\end{pmatrix} = \mathcal{R}_{\mathbf{e}}^{-1}\begin{pmatrix}|\mathbf{e}\rangle \\ |\bar{\mathbf{e}}\rangle\end{pmatrix}, \qquad \mathcal{R}_{\mathbf{e}}^{-1} \equiv \begin{pmatrix}m & n \\ ne^{-i\delta} & -me^{-i\delta}\end{pmatrix} = \mathcal{R}_{\mathbf{e}}^{\dagger} \tag{2.13}$$

In the "equatorial" plane $\chi = \pi/2$, we may have the same special cases as (2.4) – (2.7):

$$\begin{pmatrix}|\mathbf{e}\rangle \\ |\bar{\mathbf{e}}\rangle\end{pmatrix} \to \begin{pmatrix}|\odot\rangle \\ |\otimes\rangle\end{pmatrix} \text{ when } \delta \to \begin{pmatrix}0 \\ \pi\end{pmatrix} \quad \text{and} \quad \begin{pmatrix}|\mathbf{e}\rangle \\ |\bar{\mathbf{e}}\rangle\end{pmatrix} \to \begin{pmatrix}|\rightarrow\rangle \\ |\leftarrow\rangle\end{pmatrix} \text{ when } \delta \to \begin{pmatrix}\pi/2 \\ 3\pi/2\end{pmatrix}.$$

The links (2.3) and (2.11) allow us to express $|\mathbf{s}\rangle, |\bar{\mathbf{s}}\rangle$ directly in terms of $|\mathbf{e}\rangle, |\bar{\mathbf{e}}\rangle$, that is, in the $S_{\mathbf{e}}$-basis. Combining (2.3) and (2.13) and denoting $\varphi - \delta \equiv \eta$ yields:

$$\begin{pmatrix}|\mathbf{s}\rangle \\ |\bar{\mathbf{s}}\rangle\end{pmatrix} = \mathcal{R}_{\mathbf{s}}^{\mathbf{e}}\begin{pmatrix}|\mathbf{e}\rangle \\ |\bar{\mathbf{e}}\rangle\end{pmatrix}, \qquad \mathcal{R}_{\mathbf{s}}^{\mathbf{e}} \equiv \mathcal{R}_{\mathbf{s}}\,\mathcal{R}_{\mathbf{e}}^{-1} = \begin{pmatrix}u & v \\ -e^{i\eta}v^* & e^{i\eta}u^*\end{pmatrix} \tag{2.14}$$

or

$$\begin{aligned}|\mathbf{s}\rangle &= u|\mathbf{e}\rangle + v|\bar{\mathbf{e}}\rangle \\ |\bar{\mathbf{s}}\rangle &= -e^{i\eta}\left(v^*|\mathbf{e}\rangle - u^*|\bar{\mathbf{e}}\rangle\right)\end{aligned}, \qquad \begin{aligned}u &\equiv am + bn\,e^{i\eta} \\ v &\equiv an - bm\,e^{i\eta}\end{aligned} \tag{2.15}$$

Note that transformation (2.15) cannot be generally expressed in terms of some *real* angle $\gamma$ between $\mathbf{s}$ and $\mathbf{e}$. An attempt to represent $u$, $v$ by analogy with (2.1, 10) in terms of $\gamma$ gives, according to (2.15)

$$\begin{aligned}u &= \cos\frac{\gamma}{2} = \cos\frac{\theta}{2}\cos\frac{\chi}{2} + \sin\frac{\theta}{2}\sin\frac{\chi}{2}\,e^{i\eta}, \\ v &= \sin\frac{\gamma}{2} = \cos\frac{\theta}{2}\sin\frac{\chi}{2} - \sin\frac{\theta}{2}\cos\frac{\chi}{2}\,e^{i\eta}\end{aligned} \tag{2.16}$$

We see that vectors $\mathbf{s}$ and $\mathbf{e}$ do not make a real angle with each other, except for some trivial cases like $\theta = 0$ or $\eta = 0, \pi$. This is another snag in mapping spin states from complex vector space $\mathcal{H}$ onto $V$. But the probabilities of the respective outcomes $|\mathbf{e}\rangle$ or $|\bar{\mathbf{e}}\rangle$ in measurements of state $|\mathbf{s}\rangle$ in the $S_{\mathbf{e}}$-basis are real functions of $\theta$, $\chi$ and $\eta$:

$$\begin{aligned}\mathcal{P}(\mathbf{e}) &= a^2 m^2 + b^2 n^2 + 2abmn\cos\eta, \\ \mathcal{P}(\bar{\mathbf{e}}) &= a^2 n^2 + b^2 m^2 - 2abmn\cos\eta\end{aligned} \qquad \mathcal{P}(\mathbf{e}) + \mathcal{P}(\bar{\mathbf{e}}) = 1 \tag{2.17}$$

Dependence of individual probabilities on phase difference $\eta$ between $|\mathbf{s}\rangle$ and $|\mathbf{e}\rangle$ indicates particle interference with itself similar to a single photon interference in the Mach-Zehnder interferometer. The difference between the two probabilities is also periodic function of $\eta$,

$$\Delta \mathcal{P}(\eta) \equiv \mathcal{P}(\mathbf{e}) - \mathcal{P}(\bar{\mathbf{e}}) = (a^2 - b^2)(m^2 - n^2) + 4abmn \cos \eta \qquad (2.18)$$

The amplitude of the periodic term in (2.17) gives the measure of visibility (contrast) $\mathcal{V}$ of the interference pattern. It is close to maximal when either superposition (2.1) or (2.10) is equally-weighted (either $\mathbf{s}$ or $\mathbf{e}$ lies in the equatorial plane) and becomes maximal when both of them do:

$$\mathcal{V} = \begin{cases} mn, & a = b = 1/\sqrt{2} \\ ab, & m = n = 1/\sqrt{2} \\ 1/2, & a = b = m = n = 1/\sqrt{2} \end{cases} \qquad (2.19)$$

The opposite case is observed in the limit $|\mathbf{s}\rangle \to |\uparrow\rangle$ or $|\mathbf{e}\rangle \to |\uparrow\rangle$. Then either $\varphi$ or $\delta$ and thereby $\eta$ become indeterminate, so (2.17) may seem to lose the physical meaning in that limit. But this is precluded by $b$ or $n$ approaching zero, so the interference term just disappears and both probabilities become constant regardless of $\eta$. We will then have $\mathcal{V} = 0$ with

$$\left. \begin{array}{ll} \mathcal{P}(\mathbf{e}) = m^2, & \mathcal{P}(\bar{\mathbf{e}}) = n^2, \text{ when } b = 0 \\ \mathcal{P}(\mathbf{e}) = a^2, & \mathcal{P}(\bar{\mathbf{e}}) = b^2, \text{ when } n = 0 \end{array} \right\} \qquad (2.20)$$

This is a natural feature for a state represented in or measured from its reference basis.

### 3. Spin-entangled composite states

Consider first two *disentangled* particles A and B. The corresponding qubit pair can be represented by vectors $\mathbf{s}$, $\mathbf{s}'$ with the respective azimuth angles $\varphi$, $\varphi'$. We assume their spins having equal but opposite $z$-projections. Then their polar angles $\theta$, $\theta'$ must be related by $\theta' = \pi - \theta$, while $\varphi$ and $\varphi'$, and thereby $\phi \equiv \varphi - \varphi'$ may be arbitrary (Fig. 2).

If $\phi = \pm \pi$, then $\mathbf{s}$ and $\bar{\mathbf{s}}'$ are anti-parallel (case 2a). If $\phi = 0$, both vectors have identical components onto the (*x, y*)-plane (case 2b). And Fig. 2c represents the general case. This approach involves the *individual* spin states $s$, $s_z$ of each particle (uncoupled representation). The composite states can also be visualized in the coupled representation [4, 6] describing the system by the *net* spin $S$ and its Z-projection $S_Z$ (the *net* values of observables will be denoted by italics capitals). The two representations are generally different and form different bases in $\mathcal{H}$. And even though the coupled representation gives in the considered case the same information, its geometric visualization is also different from that in Fig. 2. Instead of an individual vector for each particle, we will have one geometric object for the whole pair (Fig. 3).

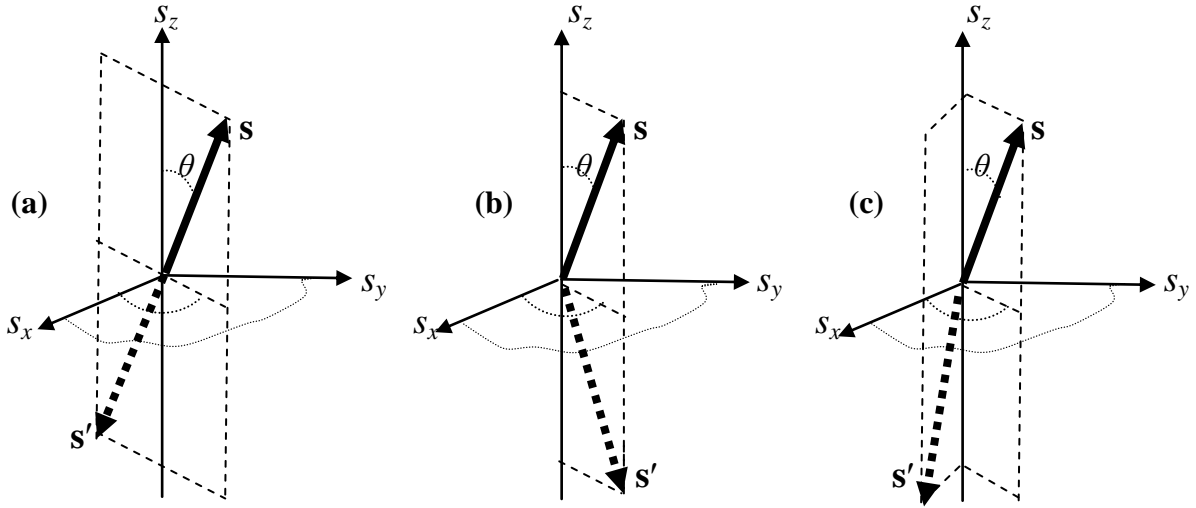

**Fig. 2**

Spin states with opposite $S_z$-components in a *disentangled* pair (not to scale). Each state is represented by the arrow, solid for A and dotted for B. Their common origin does not imply common location of the particles.

(a) The vectors are anti-parallel. If azimuth of A is $\varphi$, the azimuth of B is $\varphi' = \varphi \pm \pi$

(b) The vectors **S** and **S'**, while being opposite in the Z-dimension, have the same $\varphi$

(c) The phase difference $\phi \equiv \varphi - \varphi'$ between **S** and **S'** lies within the range $0 < |\phi| < \pi$

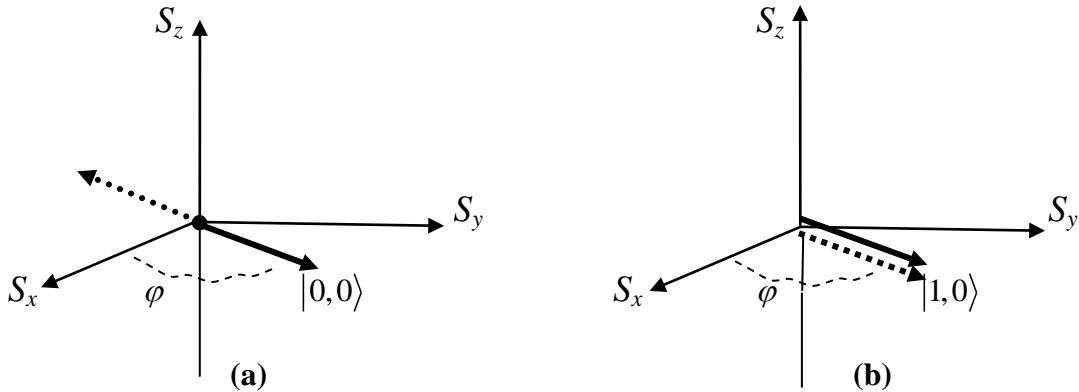

**Fig. 3**

The same as Fig. 2, but in the coupled representation

(a) $\phi = \pi$. The spin vectors of both particles before measurement are opposite. This corresponds to a singlet state $|\mathbf{S}\rangle = |S, S_z\rangle = |0,0\rangle$. Its vector image in $V$ is just the point at the origin

(b) $\phi = 0$. The individual spin vectors are opposite in the Z-dimension but have zero phase difference. The corresponding $\varphi = \varphi'$ singles out an azimuth at which the measurements in the XY plane give (+)correlated outcomes. This corresponds to a triplet $|\mathbf{S}\rangle = |S, S_z\rangle = |1,0\rangle$. It is represented as an arrow with azimuth $\varphi$ in the *XY* plane.

Both representations show that varying $\phi$ dramatically changes physical state of the system. Changing $\phi$ from $\pi$ to zero converts the net spin from 0 to 1. Any $\phi$ in between corresponds to superposition of the two. This correlation reminds the case of two interacting identical particles, when the phase difference $\Delta\phi$ between the amplitudes of their scattering into the same state determines their physical nature – whether they are bosons ($\Delta\phi = 0$) or fermions ($\Delta\phi = \pi$) [7, 8]. The underlying physics here is different (e.g., $\Delta\phi$ has only two values), but it also shows intimate connection between system's properties and the corresponding phase.

Now we turn to entangled states. Entanglement embraces much more than usually discussed composite states. It may as well involve different characteristics of a *single particle*. A spin measurement in Stern-Gerlach experiment uses entanglement between spin and momentum and thereby evolving position of a single atom in an external magnetic field. While spin is purely quantum characteristic, the position may reside in the domain of classical approximation and form the pointer states when the field is not uniform. Under used experimental conditions we infer information about spin from position to which the atomic wave packet collapses on the observation screen [4].

But here we consider a bipartite system. Specifically, we discuss *spin-entangled* particles A and B, e.g., an electron pair with some definite $S_z$. In such cases, neither particle can be imaged as a single vector like in Fig.1 or 2: since neither of them has a definite state, it is not determined which vector represents which particle. We can only treat the system analytically.

A general expression for a composite entangled state with $S_z = 0$ in the $S_z$-basis is

$$|\Psi\rangle = p|\uparrow\rangle_A |\downarrow\rangle_B + \tilde{q}|\downarrow\rangle_A |\uparrow\rangle_B \text{ with } \tilde{q} \equiv qe^{i\alpha} \text{ and } p^2 + q^2 = 1 \qquad (3.1)$$

(The used notations imply $p$ and $q$ being real positive without loss of generality). The $S_z$-measurements of $|\Psi\rangle$ will always show the two particles with *anti-parallel* spin components. This kind of correlated outcomes is frequently named anti-correlations [9 - 11]. But to avoid possible confusion with non-correlated states, we will hereafter call the composite states with opposite individual outcomes for its constituents "$(-)correlations$", and states with identical outcomes "$(+)correlations$". In case (3.1), both outcomes $|\uparrow\rangle_A |\downarrow\rangle_B$ or $|\downarrow\rangle_A |\uparrow\rangle_B$, while being physically different, are $(-)$correlations. Their relative weight is given by the ratio

$$\varepsilon \equiv p^2 / q^2 \qquad (3.2)$$

The corresponding probabilities $\mathcal{P}_p$, $\mathcal{P}_q$ expressed in terms of $\varepsilon$ are

$$\mathcal{P}_p \equiv p^2 = \frac{\varepsilon}{1+\varepsilon} \; ; \quad \mathcal{P}_q \equiv q^2 = \frac{1}{1+\varepsilon} \qquad (3.3)$$

Parameter $\varepsilon$ characterizes the degree of entanglement. State (3.1) is maximally entangled at $\varepsilon = 1$ (superposition (3.1) is equally-weighted):

$$|\Psi\rangle = \frac{1}{\sqrt{2}} \left( |\uparrow\rangle_A |\downarrow\rangle_B + e^{i\alpha} |\downarrow\rangle_A |\uparrow\rangle_B \right) \tag{3.4}$$

At weak entanglement ($\varepsilon \ll 1$ or $\varepsilon \gg 1$) one of the $(-)$ correlated outcomes, $|\uparrow\rangle_A |\downarrow\rangle_B$ or $|\downarrow\rangle_A |\uparrow\rangle_B$, becomes much more probable than the other, making the respective term in (3.1) overwhelmingly dominating and thus bringing *each* particle closer to a definite state. In the limit $\varepsilon \to 0$ or $\varepsilon \to \infty$, (3.1) reduces to only one term, the entanglement vanishes, each particle acquires its own state, but these individual states remain strictly $(-)$ correlated in the $S_Z$-basis.

How can we monitor the parameters of state (3.1)? The answer to this question depends on many factors, including physical characteristics of a pre-existing system producing (A, B)-pair. Expression (3.1) with arbitrary $p$, $\tilde{q}$ describes all possible cases with $S_Z = 0$.

Our next step will be studying the system in an arbitrary basis.

## 4. Various faces of entanglement

The spin-measurement outcomes depend on the pair's net spin. For a pure singlet *all* net spin outcomes must be zero. Accordingly, all individual results are strictly $(-)$ correlated in *any* basis – the particles must have the opposite spin components onto any axis. But state (3.1), while showing strict $(-)$ correlations in the $S_Z$-basis, may show $(+)$ *correlated* outcomes in another basis, which would be already a signature of a triplet. This reflects the fact that in QM even definite information (no entanglement) about parts of a system may be still insufficient for knowledge of the whole system [4, 6]. It is even more so when such information is indefinite like in (3.1), apart from the fact that condition $S_Z = 0$ does not define the net spin $S$.

The basis-dependence of the initially pure $(-)$ or $(+)$ correlations had been used in the ground-breaking discussions of possibility of superluminal signaling between separated locations [12-16]. In this article we formulate the quantitative criteria for state (3.1) to describe either a singlet, or triplet, or their superposition. The corresponding conditions obtain from representing the initial state in an arbitrary **e**-basis, that is, considering its behavior under basis rotation in $\mathcal{H}$.

Writing (2.12) once for A and then for B and putting into (3.1) gives after some algebra

$$|\Psi\rangle = e^{-i\delta} \left[ \tilde{f} \left( |\mathbf{e}\rangle_A |\mathbf{e}\rangle_B - |\bar{\mathbf{e}}\rangle_A |\bar{\mathbf{e}}\rangle_B \right) - \tilde{g} |\mathbf{e}\rangle_A |\bar{\mathbf{e}}\rangle_B - \tilde{h} |\bar{\mathbf{e}}\rangle_A |\mathbf{e}\rangle_B \right], \tag{4.1}$$

where

$$\tilde{f} \equiv (p + \tilde{q}) mn, \quad \tilde{g} \equiv pm^2 - \tilde{q}n^2, \quad \tilde{h} \equiv pn^2 - \tilde{q}m^2 \tag{4.2}$$

The immaterial common factor $e^{-i\delta}$ here represents the symmetry of result (4.1) with respect to rotation of **e** around the Z-axis.

Usual textbook examples of entanglement focus on cases with pure $(+)$ or $(-)$ correlations: if A and B are entangled with respect to a certain characteristic, then its measurement on A automatically determines the corresponding outcome on B. The result (4.1) shows the set of possibilities is more general. The terms with amplitudes $\tilde{g}, \tilde{h}$ form an entangled superposition of $(-)$ correlated states. But the terms with amplitude $\tilde{f}$ form a superposition of $(+)$ correlations. The $S_\mathbf{e}$- measurement will generally produce a mixture of $(-)$ correlated *and* $(+)$ correlated pairs, so *both* types of correlations can show up in an arbitrary basis. If observer Alice finds A in the $|\mathbf{e}\rangle_A$-state, her partner Bob may find B either in the $|\bar{\mathbf{e}}\rangle_B$-state or in the $|\mathbf{e}\rangle_B$-state. This is immediately seen if we rewrite (4.1) as

$$|\Psi\rangle = e^{-i\delta}\left[|\mathbf{e}\rangle_A\left(\tilde{f}|\mathbf{e}\rangle_B - \tilde{g}|\bar{\mathbf{e}}\rangle_B\right) - |\bar{\mathbf{e}}\rangle_A\left(\tilde{h}|\mathbf{e}\rangle_B + \tilde{f}|\bar{\mathbf{e}}\rangle_B\right)\right] \quad (4.3)$$

If Alice observes the outcome $|\mathbf{e}\rangle_A$ in her $S_\mathbf{e}$-measurement, then B collapses to a *superposition* $\tilde{f}|\mathbf{e}\rangle_B - \tilde{g}|\bar{\mathbf{e}}\rangle_B$ rather than just to $|\bar{\mathbf{e}}\rangle_B$, so there is a chance $|\tilde{f}|^2$ for Bob to get outcome $|\mathbf{e}\rangle_B$. And if Alice records the result $|\bar{\mathbf{e}}\rangle_A$, then B collapses to *another superposition* $\tilde{h}|\mathbf{e}\rangle_B + \tilde{f}|\bar{\mathbf{e}}\rangle_B$ instead of just $|\mathbf{e}\rangle_B$, so there is the same chance $|\tilde{f}|^2$ for Bob to find B also in state $|\bar{\mathbf{e}}\rangle_B$. In either case, only probabilistic prediction can be made for measurement on B.

Physically, expression (4.1) or (4.3) *is just state* (3.1) *written in the* $\mathbf{e}$-*basis*. The appearance of new correlations here does not mean disappearance of entanglement. Expression (4.1) remains inseparable – entanglement conserves but changes its face showing now both types of correlation. The strict $(-)$ correlation in (3.1) is not by itself a sufficient condition to describe a singlet. The emergence of $(+)$ correlations in (4.1) is a signature of the *net* spin 1. In the new basis, the system is extended into additional two dimensions of $\mathcal{H}$.

Denote the respective probabilities of 4 outcomes in (4.1) as $\mathcal{P}^+(\mathbf{e}, \mathbf{e})$, $\mathcal{P}^+(\bar{\mathbf{e}}, \bar{\mathbf{e}})$, $\mathcal{P}^-(\mathbf{e}, \bar{\mathbf{e}})$, $\mathcal{P}^-(\bar{\mathbf{e}}, \mathbf{e})$, with the first argument in parentheses standing for A and the second one for B. They are directly calculated from (4.1, 2):

$$\mathcal{P}^+(\mathbf{e}, \mathbf{e}) = \mathcal{P}^+(\bar{\mathbf{e}}, \bar{\mathbf{e}}) = |\tilde{f}|^2 = (mn)^2(1 + 2pq\cos\alpha), \quad (4.4)$$

$$\mathcal{P}^+ \equiv \mathcal{P}^+(\mathbf{e}, \mathbf{e}) + \mathcal{P}^+(\bar{\mathbf{e}}, \bar{\mathbf{e}}) = 2(mn)^2(1 + 2pq\cos\alpha); \quad (4.5)$$

and

$$\mathcal{P}^-(\mathbf{e}, \bar{\mathbf{e}}) = |\tilde{g}|^2 = (pm^2)^2 + (qn^2)^2 - 2pq(mn)^2\cos\alpha \quad (4.6)$$

$$\mathcal{P}^-(\bar{\mathbf{e}}, \mathbf{e}) = |\tilde{h}|^2 = (pn^2)^2 + (qm^2)^2 - 2pq(mn)^2\cos\alpha \quad (4.7)$$

$$\mathcal{P}^- = \mathcal{P}^-(\mathbf{e}, \bar{\mathbf{e}}) + \mathcal{P}^-(\bar{\mathbf{e}}, \mathbf{e}) = m^4 + n^4 - 4pq(mn)^2\cos\alpha \quad (4.8)$$

(It is easy to see that $\mathcal{P}_{Net} = \mathcal{P}^+ + \mathcal{P}^- = 1$). The ratio of $(+)$ correlated to $(-)$ correlated outcomes in (4.5), (4.8) is given by

$$\rho \equiv \frac{\mathcal{P}^+}{\mathcal{P}^-} = \frac{2|\tilde{f}|^2}{|\tilde{g}|^2 + |\tilde{h}|^2} \quad (4.9)$$

The only way to guarantee the zero *net* spin of the system is to require $\rho = 0$, that is to eliminate the $\tilde{f}$-terms. In view of (4.2) this gives

$$\tilde{f} \equiv (p + \tilde{q})mn = 0 \qquad (4.10)$$

Disregarding the trivial cases $m = 0$ or $n = 0$ simply taking **e** to the $s_z$-basis, we get

$$p + \tilde{q} = 0 \qquad (4.11)$$

This gives

$$\boxed{p = q = \frac{1}{\sqrt{2}} \text{ and } \alpha = \pi} \qquad (4.12)$$

thus reducing (3.4) to

$$|\Psi\rangle = \frac{1}{\sqrt{2}}\left(|\uparrow\rangle_A|\downarrow\rangle_B - |\downarrow\rangle_A|\uparrow\rangle_B\right) \qquad (4.13)$$

In this special case the initial state remains $(-)$ correlated in any basis. This is a hallmark of a bipartite singlet ($S = 0$), so *conditions* (4.12) *give the sought-for criterion.*
The opposite case – converting to strict (+) correlations under change of bases – corresponds to a pure triplet ($S = 1$) and occurs when

$$\tilde{g} = \tilde{h} = 0 \qquad (\rho^{-1} = 0) \qquad (4.14)$$

Consulting with (4.2) gives

$$\boxed{m = n = \frac{1}{\sqrt{2}} \text{ and } p = q = \frac{1}{\sqrt{2}}, \ \alpha = 0} \qquad (4.15)$$

This takes us to (3.4) with $\alpha = 0$:

$$|\Psi\rangle = \frac{1}{\sqrt{2}}\left(|\uparrow\rangle_A|\downarrow\rangle_B + |\downarrow\rangle_A|\uparrow\rangle_B\right) \qquad (4.16)$$

In this special case state $|\Psi\rangle$ posing as a singlet in the $S_z$-basis is actually a pure triplet ($S = 1$) and *conditions* (4.15) *give the criterion for it.* State (4.16) is, again, maximally entangled and $(-)$ correlated, but any $S_X$- or $S_Y$-measurements will now convert it to its opposite – rigorous $(+)$ correlations. Its initial $(-)$-correlation in (4.16) means the zero projection of the *net* spin 1 onto the Z-axis.
Conditions (4.14, 15) for pure triplet are more restrictive than (4.12) for pure singlet. Apart from requiring superposition (3.1) to be equally-weighted, it takes **e**, in view of (2.10), to the "equatorial" plane of the Bloch sphere. Only in this plane will we observe complete conversion to pure (+) correlations when measuring state (4.16). Note also that (3.2) and (4.9) describe totally different characteristics of the system. The (3.2) gives relative weights of the two composite $(-)$ correlated eigenstates in the $S_z$-basis, while (4.9) describes the ratio of $(+)$ and $(-)$ correlations in an *arbitrary* basis. Accordingly, the phase $\alpha$ is irrelevant in (3.2) while being crucial in (4.9).

On the other hand, the role of $\alpha$ here and in (3.1, 4) is similar to $\phi$ in a *disentangled* pair as visualized in Fig-s 2, 3. Changing $\alpha$ from $\pi$ to zero in (3.4) corresponds, as in that case, to the switch from a singlet to triplet state with the net spin $S$ lying in the $(X, Y)$-plane. The physical meaning of $\alpha$ is different from that of $\phi$ but its numerical values are also connected with the corresponding properties of the system.

Generally, with amplitudes $p$ and $\tilde{q}$ being arbitrary within normalization condition, expression (3.1) describes a superposition of singlet and triplet states with restriction $S_z = 0$. The weights of these states are determined by the respective amplitudes in (4.1) or (4.3). According to (4.9), they are equally-weighted when

$$\mathcal{P}^+ = \mathcal{P}^- \quad \text{or} \quad 2|\tilde{f}|^2 = |\tilde{g}|^2 + |\tilde{h}|^2 = \frac{1}{2} \qquad (4.17)$$

For more detailed analytical evaluation, express (4.17) in terms of $p, \tilde{q}, m, n$:

$$2(mn)^2 + 4pq(mn)^2 \cos\alpha = m^4 + n^4 - 4pq(mn)^2 \cos\alpha = \frac{1}{2} \qquad (4.18)$$

For each given triad $p, q, \alpha$ we get the solutions of (4.18) for $m, n$:

$$\boxed{m^2 = \frac{1}{2}\left(1 \pm \sqrt{\frac{2pq\cos\alpha}{1+2pq\cos\alpha}}\right); \quad n^2 = \frac{1}{2}\left(1 \mp \sqrt{\frac{2pq\cos\alpha}{1+2pq\cos\alpha}}\right)} \qquad (4.19)$$

*This criterion selects the basis in which the $(+)$ and $(-)$ correlations become equally-weighted.* An interesting result here is that, unless (3.1) gets disentangled, conditions (4.17) can be met (Eq. (4.19) has real solutions) only for the region $0 \le \alpha \le \pi/2$ or $3\pi/2 \le \alpha \le 2\pi$. Outside these regions, $(+)$ and $(-)$ correlations may coexist but are not equally-weighted. Within these regions, the resulting mix of the respective measurement outcomes becomes totally random, so the strict $(-)$ correlation of (3.1) is completely lost. *State* (3.1) *apparently posing as a singlet, is generally a superposition of the singlet and triplet states.*

An important result of the above analysis is periodic dependence of all probabilities in (4.4-8) on phase angle $\alpha$. As in (2.17), it is a signature of interference, but this time it is two-particle interference, and variable $\alpha$ is different from $\eta$ in (2.17). The amplitude of the periodic term determines, as in (2.17), the contrast of the interference pattern:

$$\mathcal{V} = 4pq(mn)^2 \qquad (4.18)$$

It is maximal when superposition (3.1) is equally-weighted and **e** lies in the equatorial plane.

The most important feature of the whole phenomenon is vanishing of one-particle interference in all considered entangled states. This becomes evident if we find from (4.4-8) the net probability, say, for A to collapse to a state $|e\rangle$:

$$\mathcal{P}_A(\mathbf{e}) = \mathcal{P}^+(\mathbf{e}, \mathbf{e}) + \mathcal{P}^-(\mathbf{e}, \bar{\mathbf{e}}) = (mn)^2 + (pm^2)^2 + (qn^2)^2 \qquad (4.19)$$

Similarly,

$$\mathcal{P}_B(\mathbf{e}) = \mathcal{P}^+(\mathbf{e}, \mathbf{e}) + \mathcal{P}^-(\bar{\mathbf{e}}, \mathbf{e}) = (mn)^2 + (pn^2)^2 + (qm^2)^2 \qquad (4.20)$$

Probabilities of *local* measurement outcomes here do not depend on $\alpha$, whereas probabilities of *nonlocal* outcomes in (4.4-8) are periodic functions of $\alpha$. The entanglement "steals" the coherence from its local constituents and transfers it to the whole pair. The same result will be obtained in the next section for (+)correlated entanglement. This shows that the described effect is a general characteristic of entangled composite states. Since each nonlocal state is a direct product (e.g., $|\uparrow\rangle_A |\downarrow\rangle_B$) of local states, it is natural that the coherence must shift "up" to the nonlocal level.

## 5. Entangled states with $S_z = 1$

For completeness, we need to consider also a state which is $(+)$ correlated in the reference basis:

$$|\Phi\rangle = p|\uparrow\rangle_A |\uparrow\rangle_B + \tilde{q}|\downarrow\rangle_A |\downarrow\rangle_B \qquad (5.1)$$

Generally, $|\Phi\rangle$ may be, like $|\Psi\rangle$ in (3.1), a superposition of singlet and triplet states, but it behaves differently under rotations in $\mathcal{H}$. Now the same procedure as one performed for state (3.1) leads to

$$|\Phi\rangle = p\left(m|e\rangle_A + n|\bar{e}\rangle_A\right)\left(m|e\rangle_B + n|\bar{e}\rangle_B\right) + \tilde{q}e^{-2i\delta}\left(n|e\rangle_A - m|\bar{e}\rangle_A\right)\left(n|e\rangle_B - m|\bar{e}\rangle_B\right) \quad (5.2)$$

Rearranging gives

$$|\Phi\rangle = \tilde{F}|e\rangle_A |e\rangle_B + \tilde{G}|\bar{e}\rangle_A |\bar{e}\rangle_B + \tilde{H}\left(|e\rangle_A |\bar{e}\rangle_B + |\bar{e}\rangle_A |e\rangle_B\right), \quad (5.3)$$

where

$$\tilde{F} \equiv pm^2 + qe^{i\xi}n^2, \quad \tilde{G} \equiv pn^2 + qe^{i\xi}m^2, \quad \tilde{H} \equiv \left(p - qe^{i\xi}\right)mn, \qquad (5.4)$$

and

$$\xi \equiv \alpha - 2\delta \qquad (5.5)$$

Unlike (4.1, 3), the term $e^{-i\delta}$ cannot be factored out here. The reason is that a triplet may have a nonzero projection onto the equatorial plane, similar to that in Fig. 3b. This singles out the corresponding azimuth, which excludes cylindrical symmetry of the system, so the $s_e$-measurement outcomes become sensitive to $\delta$ through the phase (5.5).

The corresponding probabilities are

$$\mathcal{P}^+(\mathbf{e}, \mathbf{e}) = |\tilde{F}|^2 = p^2 m^4 + q^2 n^4 + 2pq(mn)^2 \cos\xi, \quad (5.6)$$

$$\mathcal{P}^+(\bar{\mathbf{e}}, \bar{\mathbf{e}}) = |\tilde{G}|^2 = p^2 n^4 + q^2 m^4 + 2pq(mn)^2 \cos\xi; \quad (5.7)$$

$$\mathcal{P}^+ \equiv \mathcal{P}^+(\mathbf{e}, \mathbf{e}) + \mathcal{P}^+(\bar{\mathbf{e}}, \bar{\mathbf{e}}) = m^4 + n^4 + 4pq(mn)^2 \cos\xi \quad (5.8)$$

$$\mathcal{P}^-(\mathbf{e}, \bar{\mathbf{e}}) = \mathcal{P}^-(\bar{\mathbf{e}}, \mathbf{e}) = |\tilde{H}|^2 = (mn)^2(1 - 2pq\cos\xi) \quad (5.9)$$

$$\mathcal{P}^- = \mathcal{P}^-(\mathbf{e}, \bar{\mathbf{e}}) + \mathcal{P}^-(\bar{\mathbf{e}}, \mathbf{e}) = 2(mn)^2(1 - 2pq\cos\xi) \quad (5.10)$$

$$\mathcal{P}_{Net} = \mathcal{P}^+ + \mathcal{P}^- = 1. \quad (5.11)$$

State (5.1) remains (+)correlated if

$$\tilde{H} = 0 \quad (5.12)$$

This condition gives, in view of (5.4, 5):

$$\boxed{p = q, \quad \delta = \alpha/2}, \quad (5.13)$$

which reduces (5.1) to

$$|\Phi\rangle = \frac{1}{\sqrt{2}}\left(|\uparrow\rangle_A |\uparrow\rangle_B + e^{i\alpha}|\downarrow\rangle_A |\downarrow\rangle_B\right) \quad (5.14)$$

Alternatively, one of the amplitudes $m, n$ may be zero. This possibility is trivial – it just takes basis $(|\mathbf{e}\rangle, |\bar{\mathbf{e}}\rangle)$ to $(|\uparrow\rangle, |\downarrow\rangle)$.

The conditions for the initial (+)correlations to be converted to pure $(-)$ correlations are

$$\tilde{F} = 0, \quad \tilde{G} = 0 \quad (5.15)$$

which gives

$$\boxed{\delta = \frac{\alpha - \pi}{2}, \quad p = q, \quad m = n} \quad (5.16)$$

Physically, this requires superposition (5.1) to be equally-weighted *and* $\mathbf{e}$ to lie in the equatorial plane with $\delta$ satisfying (5.16). The local probabilities, say, for A are:

$$\mathcal{P}_A(\mathbf{e}) \equiv \mathcal{P}^+(\mathbf{e}, \mathbf{e}) + \mathcal{P}^-(\mathbf{e}, \bar{\mathbf{e}}) = p^2 m^4 + q^2 n^4 + (mn)^2 \quad (5.17)$$

$$\mathcal{P}_A(\bar{\mathbf{e}}) \equiv \mathcal{P}^-(\bar{\mathbf{e}}, \mathbf{e}) + \mathcal{P}^+(\bar{\mathbf{e}}, \bar{\mathbf{e}}) = p^2 n^4 + q^2 m^4 + (mn)^2 \quad (5.18)$$

$$\mathcal{P}_A \equiv \mathcal{P}_A(\mathbf{e}) + \mathcal{P}_A(\bar{\mathbf{e}}) = 1 \quad (5.19)$$

and similar expressions for B.

We see the same trend as in the previous section. The local probabilities are phase-independent, whereas nonlocal ones are periodic functions of $\xi$. Again, the coherence is shifted by entanglement from local to nonlocal states.

# 6. Mixed bases

Consider now measuring A and B in different bases: for instance, use $(|\mathbf{e}\rangle, |\bar{\mathbf{e}}\rangle)$ for A and some other basis $(|\mathbf{e}'\rangle, |\bar{\mathbf{e}}'\rangle)$ for B. The relationship between $(|\mathbf{e}'\rangle, |\bar{\mathbf{e}}'\rangle)$ and $(|\uparrow\rangle, |\downarrow\rangle)$ is analytically identical to (2.10, 12) but has different numerical values:

$$\left.\begin{array}{l}|\mathbf{e}'\rangle = m'|\uparrow\rangle + n'e^{i\delta'}|\downarrow\rangle \\ |\bar{\mathbf{e}}'\rangle = n'|\uparrow\rangle - m'e^{i\delta'}|\downarrow\rangle\end{array}\right\}, \qquad \left.\begin{array}{l}|\uparrow\rangle = m'|\mathbf{e}'\rangle + n'|\bar{\mathbf{e}}'\rangle \\ |\downarrow\rangle = e^{-i\delta'}\left(n'|\mathbf{e}'\rangle - m'|\bar{\mathbf{e}}'\rangle\right)\end{array}\right\}, \qquad (6.1)$$

with

$$m' \equiv \cos\frac{\chi'}{2}, \quad n' \equiv \sin\frac{\chi'}{2}, \quad \delta \to \delta' \qquad (6.2)$$

State (3.1) will be described in $(|\mathbf{e}'\rangle, |\bar{\mathbf{e}}'\rangle)$ basis by the same expressions as (4.1, 2), with all respective characteristics primed:

$$|\Psi\rangle = e^{-i\delta'}\left[\tilde{f}'\left(|\mathbf{e}'\rangle_A|\mathbf{e}'\rangle_B - |\bar{\mathbf{e}}'\rangle_A|\bar{\mathbf{e}}'\rangle_B\right) - \tilde{g}'|\mathbf{e}'\rangle_A|\bar{\mathbf{e}}'\rangle_B - \tilde{h}'|\bar{\mathbf{e}}'\rangle_A|\mathbf{e}'\rangle_B\right], \qquad (5.3)$$

$$\tilde{f}' \equiv (p+\tilde{q})m'n', \quad \tilde{g}' \equiv pm'^2 - \tilde{q}n'^2, \quad \tilde{h}' \equiv pn'^2 - \tilde{q}m'^2 \qquad (5.4)$$

Let now Alice measure her particle in the $(|\mathbf{e}\rangle, |\bar{\mathbf{e}}\rangle)$-basis while Bob chooses $(|\mathbf{e}'\rangle, |\bar{\mathbf{e}}'\rangle)$-basis. Then, in order to express the result in terms of all respective eigenstates, we must use (2.12) for the A-states, and its primed version for the B-states:

$$\left.\begin{array}{l}|\uparrow\rangle_A = m|\mathbf{e}\rangle_A + n|\bar{\mathbf{e}}\rangle_A \\ |\downarrow\rangle_A = e^{-i\delta}\left(n|\mathbf{e}\rangle_A - m|\bar{\mathbf{e}}\rangle_A\right)\end{array}\right\}, \qquad \left.\begin{array}{l}|\uparrow\rangle_B = m'|\mathbf{e}'\rangle_B + n'|\bar{\mathbf{e}}'\rangle_B \\ |\downarrow\rangle_B = e^{-i\delta'}\left(n'|\mathbf{e}'\rangle_B - m'|\bar{\mathbf{e}}'\rangle_B\right)\end{array}\right\} \qquad (6.5)$$

Putting this into (3.1) gives:

$$|\Psi\rangle = \tilde{\mu}|\mathbf{e}\rangle_A|\mathbf{e}'\rangle_B + \tilde{\nu}|\bar{\mathbf{e}}\rangle_A|\bar{\mathbf{e}}'\rangle_B + \tilde{\sigma}|\mathbf{e}\rangle_A|\bar{\mathbf{e}}'\rangle_B + \tilde{\tau}|\bar{\mathbf{e}}\rangle_A|\mathbf{e}'\rangle_B \qquad (6.6)$$

with

$$\left.\begin{array}{l}\tilde{\mu} \equiv pe^{-i\delta'}mn' + qe^{i(\alpha-\delta)}nm', \quad \tilde{\nu} \equiv pe^{-i\delta'}nm' + qe^{i(\alpha-\delta)}mn' \\ \tilde{\sigma} \equiv pe^{-i\delta'}mm' - qe^{i(\alpha-\delta)}nn', \quad \tilde{\tau} \equiv pe^{-i\delta'}nn' - qe^{i(\alpha-\delta)}mm'\end{array}\right\}, \qquad (6.7)$$

Expression (6.6) also shows 4 possible outcomes, as in case (4.1). If $\mathbf{e}$ resides in the Northern hemisphere of Fig. 1, then $\bar{\mathbf{e}}$ is mirrored into the Southern hemisphere. If they both lie in the equatorial plane, we assign the range $0 \leq \delta < \pi$ for $\mathbf{e}$ and $\pi \leq \delta < 2\pi$ for $\bar{\mathbf{e}}$. And the same arrangement will be used for $\mathbf{e}'$, $\bar{\mathbf{e}}'$. This allows us to broaden the definition of correlations to the cases of measurements in different bases, when the measurement outcomes for A and B are not exactly identical or exactly opposite. Namely, we will define the outcomes $|\mathbf{e}\rangle_A |\mathbf{e}'\rangle_B$ and $|\bar{\mathbf{e}}\rangle_A |\bar{\mathbf{e}}'\rangle_B$ as (+)correlated, whereas $|\mathbf{e}\rangle_A |\bar{\mathbf{e}}'\rangle_B$ and $|\bar{\mathbf{e}}\rangle_A |\mathbf{e}'\rangle_B$ as (−) correlated. This makes sense since, e.g., $\mathbf{e}'$ will be closer to $\mathbf{e}$ than to $\bar{\mathbf{e}}$. Then we can say that expression (6.6) describes a system with *generalized* (+) and (−) correlation outcomes. The corresponding probabilities obtain from (6.7) as

$$\mathcal{P}^+(\mathbf{e}, \mathbf{e}') = |\tilde{\mu}|^2 = p^2(mn')^2 + q^2(nm')^2 + 2pqmnm'n'\cos\zeta \qquad (5.8)$$

$$\mathcal{P}^+(\bar{\mathbf{e}}, \bar{\mathbf{e}}') = |\tilde{\nu}|^2 = p^2(nm')^2 + q^2(mn')^2 + 2pqmnm'n'\cos\zeta \qquad (5.9)$$

$$\mathcal{P}^+ \equiv \mathcal{P}^+(\mathbf{e}, \mathbf{e}') + \mathcal{P}^+(\bar{\mathbf{e}}, \bar{\mathbf{e}}') = (mn')^2 + (nm')^2 + 4pqmnm'n'\cos\zeta; \qquad (6.10)$$

and

$$\mathcal{P}^-(\mathbf{e}, \bar{\mathbf{e}}') = |\tilde{\sigma}|^2 = p^2(mm')^2 + q^2(nn')^2 - 2pqmnm'n'\cos\zeta \qquad (6.11)$$

$$\mathcal{P}^-(\bar{\mathbf{e}}, \mathbf{e}') = |\tilde{\tau}|^2 = p^2(nn')^2 + q^2(mm')^2 - 2pqmnm'n'\cos\zeta \qquad (6.12)$$

$$\mathcal{P}^- = \mathcal{P}^-(\mathbf{e}, \bar{\mathbf{e}}') + \mathcal{P}^-(\bar{\mathbf{e}}, \mathbf{e}') = (mm')^2 + (nn')^2 - 4pqmnm'n'\cos\zeta \qquad (6.13)$$

Here

$$\zeta \equiv \alpha - \delta + \delta', \qquad (6.14)$$

and $\mathcal{P} = \mathcal{P}^+ + \mathcal{P}^- = 1$. The local probability for A to be found in state $|\mathbf{e}\rangle_A$ regardless of the outcome for B is

$$\mathcal{P}_A(\mathbf{e}) \equiv \mathcal{P}^+(\mathbf{e}, \mathbf{e}') + \mathcal{P}^-(\mathbf{e}, \bar{\mathbf{e}}') = p^2 m^2 + q^2 n^2 \qquad (6.15)$$

For state $|\bar{\mathbf{e}}\rangle_A$ we have

$$\mathcal{P}_A(\bar{\mathbf{e}}) \equiv \mathcal{P}^+(\bar{\mathbf{e}}, \bar{\mathbf{e}}') + \mathcal{P}^-(\bar{\mathbf{e}}, \mathbf{e}') = p^2 n^2 + q^2 m^2 \qquad (6.16)$$

Similar expressions can be obtained for B. We see that in this general case, all nonlocal probabilities are, again, periodic functions of phase $\zeta$, while the local probabilities are constants depending only on *p* and *m*. As before, the coherence is transferred from individual particles to nonlocal entangled states. The same results also obtain if we measure, instead of (3.1), state (5.1) in the mixed basis.

Note that each initial composite state considered above describes correlations of *the same* sign - (−) in (3.1) or (+) in (5.1). We will say that either case shows a *pure correlation*. We saw that such correlation loses its purity under change of basis, except for some special cases like (4.11), whereas entanglement survives any rotations in $\mathcal{H}$. On the other hand, a pure correlation is, unlike entanglement, "robust" under change of superposition amplitudes *in a fixed basis* and conserves even when such change leads to disentanglement. Changing a basis is purely geometric transformation. Changing superposition amplitudes *in a fixed basis* is a physical change of quantum state.

If we switch from $S_z$ to another basis, the amplitudes will accordingly transform by rules (2.15) (this is not a physical change!), which may lead to emergence of the additional composite states as described by (4.1) or (5.2). This explains why a pure correlation becomes mixed.

A single qubit in superposition (2.1) is in both of its eigenstates at once. Similarly, a qubit pair in state (3.1) is in its two *composite eigenstates* at once. Accordingly, a nonlocal system can interfere with itself the same way as does a single qubit. This becomes obvious if we denote each composite state as a single ket using the same rule as in (4.4-8), e.g., $|\uparrow\rangle_A |\downarrow\rangle_B \equiv |\uparrow\downarrow\rangle$, $|\downarrow\rangle_A |\uparrow\rangle_B \equiv |\downarrow\uparrow\rangle$. Then state $|\Psi\rangle$ in (3.1), (4.1) can be rewritten as

$$|\Psi\rangle = \begin{cases} p|\uparrow\downarrow\rangle + \tilde{q}|\downarrow\uparrow\rangle, & S_z\text{-basis} \quad (6.17\ a) \\ e^{-i\delta}\left[\tilde{f}(|\mathbf{ee}\rangle - |\mathbf{\bar{e}\bar{e}}\rangle) - \tilde{g}|\mathbf{e\bar{e}}\rangle - \tilde{h}|\mathbf{\bar{e}e}\rangle\right], & S_\mathbf{e}\text{-basis} \quad (6.17\ b) \end{cases}$$

Each ket here is just one of the 4 eigenstates of our bipartite. Only 2 of them are present in (6.17a), while all 4 – in (6.17b). Generally, the mathematical structure of entangled superposition (6.17) is the same as in simple superposition (2.1). The only distinction is that the eigenstates now are nonlocal and $\mathcal{H}$-space is 4-dimentional. Generalizing a very useful term "photonic atom" (used also in [10, 11]), we can call entangled system (6.17) the "*nonlocal atom*". Such atom interferes with itself as does a single particle in (2.1) with the only distinction that the number of periodic terms in the interference pattern may reach 4.

## 7. Conclusions

As is well known, entanglement is an extremely fragile state of a system – it is easily destroyed (decohered) even by very weak perturbations (see, e.g., [20-26]). At the same time it is, in contrast with pure correlations, invariant under rotations in $\mathcal{H}$ – the whole system, while changing in number of superposed eigenstates, remains entangled. In our jargon, the system may show different "faces" in different bases – e.g., converting from $(+)$ or $(-)$ correlated state to its opposite or to their superposition. There are exceptions when initial pure correlation remains invariant – case (4.13) for initially $(-)$ correlated spin-entangled state and case (5.14) with condition (5.13) for initially $(+)$correlated state. Generally, an initial state with pure correlation converts to a superposition of differently correlated states, which may be equally-weighted under some specific conditions. All these changes can be reversed under inversed rotations.

There are analytical criteria for each type of behavior of entangled states, which have been formulated in the article.

Generally, entanglement and correlations are different concepts describing different characteristics of a system. The term "correlations" embraces a larger set of systems than entanglement. Two pure subsets of pairs – one $|\uparrow\rangle_A |\downarrow\rangle_B$ and the other $|\downarrow\rangle_A |\uparrow\rangle_B$ formed after $S_z$-measurement on state (3.1) – are already disentangled, but they all remain $(-)$ correlated. Moreover, correlated systems may be purely classical, whereas entanglement is exclusively quantum property. Therefore we must be careful when discussing role of correlations in entangled superposition.

The most important result of this work is that nonlocal probabilities are periodic functions of the phase difference between superposed states. Entanglement destroys periodic pattern for each separate particle and transfers coherence from single events to combined outcomes. Such effect of coherence transfer was experimentally demonstrated in 1990 for the equally-weighted superposition of momentum-entangled photons (RTO experiments [9 – 11, 18, 19]). Thus, presented analysis reveals the analogy between the interference patterns in totally different physical systems – momentum-entangled bosons and spin-entangled fermions. This shows that we are dealing with a very general phenomenon which might be also generalized back to an *arbitrary*, rather than just equally-weighted, superposition of coupled bosons. Altogether, we can predict some similar features in behavior of spin-entangled electrons and momentum-entangled photons under measurements in the respective bases. This topic will be addressed in another article.

## Acknowledgements

A want to thank Art Hobson, Emeritus Professor of Physics, University of Arkansas for inspiring discussions, and Anwar Sheikh, Colorado Mesa University, for helpful comments.

*Physics Today*, 44, pp. 36–44 (1991)
21. M. Brune a.o., Observing the Progressive Decoherence of the "Meter" in a Quantum
    Measurement, *Phys. Rev. Lett.*, **77** (24), 1996
23. D. Bacon, *"Decoherence, control, and symmetry in quantum computers"*
    arXiv:quant-ph/0105127
24. W. H. Zurek, Decoherence, einselection, and the quantum origins of the classical,
    *Rev. of Mod. Phys.,* **75,** 715. (2003)
25. M. Schlosshauer, "Decoherence, the measurement problem, and interpretations of quantum
    mechanics", *Rev. of Mod. Phys.* **76** (4), 1267–1305 (2005);
26. M. Schlosshauer, *Decoherence and the Quantum-to-Classical Transition*,
    Springer, Berlin (2007)19